\documentclass{PoS}
\title{Approximate forms of the density of states}

\ShortTitle{Approximate forms of the density of state}

\author{A. Denbleyker, Daping Du, Yuzhi Liu, and  \speaker{Y. Meurice}\\
Department of Physics and Astronomy, The University of Iowa,
Iowa City, Iowa 52242, USA\\
E-mail: \email{alan-denbleyker@uiowa.edu}\\
E-mail: \email{daping-du@uiowa.edu}\\
E-mail: \email{yuzhi-liu@uiowa.edu}\\
E-mail:\email{yannick-meurice@uiowa.edu}}
\author{A. Velytsky\\
EFI, University of Chicago, 5640 S. Ellis Ave., Chicago, IL 60637
and Argonne National Laboratory, 9700 Cass Ave., Argonne, IL 60439\\
E-mail:\email{vel@uchicago.edu} }

\abstract{We compare MC calculations of the density of states in SU(2) pure gauge theory with the weak and strong 
coupling expansions. Surprisingly, the range of validity of the two approximations overlap significantly, however the 
large order behavior of both expansions appear to be similar to the corresponding expansions of the plaquette.
We discuss the implications for the calculation of the Fisher's zeros of the partition function. }

\FullConference{The XXVI International Symposium on Lattice Field Theory\\
		 July 14-19 2008\\
		 Williamsburg, Virginia, USA}

\begin{document}

\section{Introduction}
Understanding the large distance behavior of asymptotically free gauge theories in terms 
of the weakly coupled short distance degrees of freedom is a major challenge for theoretical physics. 
In  pure gauge theory with the standard Wilson's action, the available numerical data on $L^4$ lattices indicates that there is no phase transition for $SU(2)$ or $SU(3)$ and the theory should be in the 
confining phase for all values of the coupling. 
Convincing arguments have 
been given \cite{tomboulis07,tomboulis07b} in favor of the smoothness of the renormalization group flows between the two fixed points 
corresponding to the two limits. 
This suggests that it is possible to match 
the weak coupling and the strong coupling expansions of the lattice formulation. 
However, if we consider the two expansions, for instance for the average $SU(2)$ plaquette as a function of $\beta=4/g^2$,  there is a 
crossover region 
(approximately $1.5<\beta<2.5$) where none of the two expansions seem to work.
This situation can probably be explained in terms of singularities in the complex $\beta$ plane \cite{kogut80b,third} 
that at this point are not completely understood. In these proceedings, we discuss the weak and strong coupling expansions of the 
density of states for $SU(2)$ and compare them to Monte Carlo calculations. The density of states is the inverse Laplace (or Borel) transform 
of the partition function. Its logarithm can be interpreted as a "color entropy". This is discussed  in section \ref{sec:density} where the basic concepts are defined. 

For the one plaquette model, the density of states is a function that has better convergence properties than the partition function \cite{plaquette}. 
This is explained in section \ref{sec:one}. We would like to know if this property 
persists on $L^4$ lattices. 
The comparison between weak and strong expansions and numerical calculations of 
the density of states for a $6^4$ lattice
are summarized in section \ref{sec:approximate}.  More details can be found in  \cite{denbleyker:054503}.

Knowing the density of states, we can calculate the partition function and its derivatives for any real or complex value of $\beta$.  In particular, it can be used to determine the Fisher's 
zeros of the partition function \cite{alves90b,alves91}. Locating these zeros in the complex $\beta$ plane and their volume dependence is important to understand the large order behavior of the weak coupling expansion \cite{ third,npp,quasig,lat07} at zero temperature and the nature of the finite temperature transition \cite{janke04}. Related questions have also been discussed in a poster presented at the same conference \cite{poster}.

\section{The density of states}
\label{sec:density}
\def\mn{\mathcal{N}_p}
In the following, we focus on a  $SU(2)$ gauge theory with Wilson's action on a  $L^4$ lattice and periodic 
boundary conditions. We denote the number of plaquettes
$\mn=6\times L^4$. The partition function 
$Z(\beta)$ is the Laplace transform of $n(S)$, the density of states:
\begin{equation}
Z(\beta) =\int_0^{2\mn}dS\  n(S)\ {\rm e}^{-\beta S}\ ,
\label{eq:intds}
 \end{equation}
 with
 \begin{equation}
n(S)=\prod_{l}\int dU_l \delta(S-\sum_{p}(1-(1/N)Re Tr(U_p)))
\end{equation}
 We can interpret 
ln($n(S)$) as a "color entropy" (extensive). 
For cubic lattices with  an even number of sites in each direction and a gauge group that  contains $-1$,  it is possible to change $\beta Re TrU_p$ into $-\beta Re TrU_p$ by a  change of variables $U_l\rightarrow -U_l$ 
on a set of links such that for any  plaquette,  exactly one link of the set belongs to that plaquette \cite{gluodyn04}. This implies
\begin{equation}
Z(-\beta)={\rm e}^{2\beta\mathcal{N}_p}Z(\beta)
	\label{eq:su2id}
\end{equation}
and consequently
\begin{equation}
n(2\mn -S)=n(S)
\label{eq:dual}
\end{equation}
Thanks to this symmetry, we only need to know $n(S)$ for $0\leq S\leq \mn$ .
Note that $<S>=\mn$ means $<TrU_p>$=0.

We define 
\begin{equation}
f(x,\mn)\equiv ln( n(x\mn,\mn))/\mn\  .
\end{equation}
The symmetry (\ref{eq:dual}) implies that 
\begin{equation}
f(x,\mn)=f(2-x,\mn)
\end{equation}
The existence of the infinite volume limit requires that 
\begin{equation}
lim_{\mn \rightarrow \infty} f(x,\mn) = f(x)\ ,
\end{equation}
with $f(x)$ volume independent. In the same limit, the integral ( \ref{eq:intds}) can be 
evaluated by the saddle point method. The maximization of the integrand requires 
\begin{equation}
f'(x)=\beta \  .
\label{eq:saddle}
\end{equation}

\section{The one plaquette case}
\label{sec:one}
In the case of the one plaquette model,  the density of state simply follows from the explicit form of the Haar measure:
 \begin{equation}
 n_{1 pl.}(S)=\frac{2}{\pi}\sqrt{S(2-S)}
 \end{equation} 
 At leading order, the large $\beta$ behavior of the partition function is determined by the behavior of $n(S)$ near $S=0$.  
The fact that $n(S)\propto \sqrt{S}$ for small $S$ implies 
$Z\propto \beta^{-3/2}$ for large $\beta$. The 
$1/\beta$ corrections can be 
calculated by expanding the remaining factor $\sqrt{2-S}$ 
in powers of $S$.  One then sees that a 
series with finite radius of convergence becomes an asymptotic series if we integrate 
over $S$ from 0 to $\infty$ (instead of 0 to 2). In addition, the large order behavior of 
the asymptotic series is determined by the non-analyticity of $n_{1pl.}(S)$ at the 
maximal value of $S$ (2 in this case). 

These properties are in agreement with the general idea that the large order behavior 
of the weak coupling expansion is determined by the behavior at small negative 
coupling \cite{bender69,leguillou90}.
In the present case, small negative $g^2$ means that $\beta$ is very negative. 
In this limit, the largest possible values of $S$ dominate the integral (in agreement 
with what we explained above). 
It would be interesting to understand if this property persists on $L^4$ lattices. 
Unfortunately, numerical values of the weak coupling expansion of the plaquette are 
not available for $SU(2)$ and we will have to rely on a model proposed in \cite{npp}.

\section{Approximate forms of $n(S)$}
\label{sec:approximate}

Numerical calculations of $n(S)$ can be obtained by patching plaquette distributions 
multiplied by the inverse Boltzmann weight at various  
values of $\beta$. In  
\cite{denbleyker:054503}
we presented 
numerical data for $L^4$ lattices  with $L=\ 4,\ 6$ and 8. For these values of $L$,  
finite volume effects are not too large and plaquette distributions are broad enough 
to allow a reasonably smooth patching.  The volume dependence is resolvable only for small values of $S$ where a behavior $ln(S)/V$ 
is observed for $ln(n(S))$. The coefficient of the singularity was calculated to be 
$(3/4)-(5/12)L^{-4}$ in reasonably good agreement with the numerical data. 

\begin{figure}[b]
\begin{center}
\includegraphics[width=4.in]{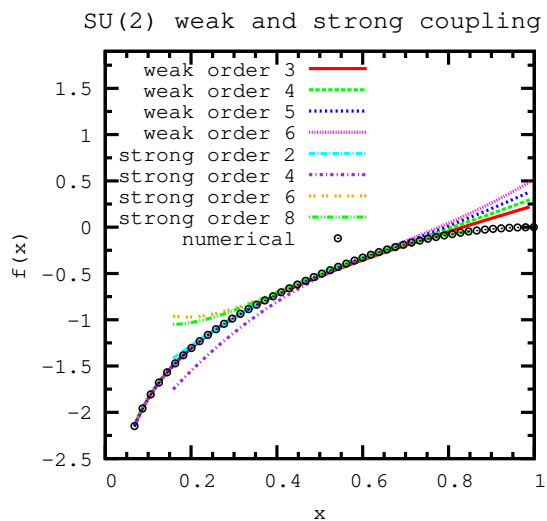}
\caption{\label{fig:wall}  Weak and strong coupling expansion of $f$ at a few intermediate orders.} 
\end{center}
\end{figure}

The numerical results for $f(x)$ were compared with expansions that can be obtained from 
the strong and weak coupling 
expansions of the average plaquette. Intermediate orders in these expansions show 
a good overlap for values of $S$ that correspond to the crossover (see Fig. \ref{fig:wall}). The convergence of the new series can be related empirically to those of the series 
for the average plaquette.  
The general picture that was obtained by trying with known series is that the converted 
series inherits the asymptotic behavior of the original series. The conversion of the 
series is performed using the saddle point equation. 
For the strong coupling, we expand about $x=1$ (we remind that $x=S/\mn$,  see section 2). Graphs of the accuracy of the expansion at successive orders, show a crossing characteristic of a finite radius of convergence near $x=0.5$. This is consistent with 
a crossing near $\beta\simeq 2$ for the plaquette (for $\beta=2$, the average plaquette 
is about 0.47).  For the weak coupling, we expand about $x=0$. Accuracy graphs show 
consistent improvement as the order increases (with possible saturation) when  $x<0.4$ for $f(x)$ and $\beta>3$ 
for the plaquette. However it should be kept in mind that the large order of the series 
for the plaquette has been modeled rather than calculated explictly. For details and graphs see \cite{denbleyker:054503}. 

The weak coupling expansion determines the logarithmic 
singularities of $ln(n(S))$ at both boundaries. When these singularities are subtracted 
we obtain a bell-shaped function that can be approximated by Legendre or Chebyshev polynomials. Empirically, the determination of the expansion coefficients based on the  discrete 
orthogonality of the Chebyshev polynomials (rather then numerical interpolation followed 
by numerical integration) seems the most stable.

\section{Calculation of Fisher's zeros}
\label{sec:zeros}

One motivation for this work is to improve our ability to determine the zeros of the partition function in the complex $\beta$ plane. For reference, it is useful to 
understand the limitation of the reweighting MC method. 
In order to estimate the errors in the location of the complex zeros, we considered the 
changes in the location of the zeros of the real and imaginary part due to statistical fluctuations.
We considered 200 sets of 40,000  values of $S$ picked at random out of the large sample of values
(bootstraps) generated for $\beta=2.225$. For each of the 200 sets, we calculated the zeros of the real part 
on a small grid with typical distance between neighbooring points of the order of $10^{-3}$. Using this procedure, 
383895 zeros of the real part were found. We then studied the distribution of these zeros 
using a 200 by 200 grid in the $\beta$ complex plane. The results are shown in Fig.  2.
In this contour plot, the outer contours go through the bins that have 20 zeros, the first inner 
contours correspond to 60 zeros, the next to 100 zeros etc..
The circle of confidence \cite{alves91} in the Gaussian approximation for 40,000 independent configurations as well as another estimate (red boxes in Fig. 2) of the region of confidence discussed in \cite{quasig} are  
shown on this graph for reference. It is clear that as we get closer to the boundary of the region of confidence, 
the distributions get wider. 

It is easier to look at horizontal sections of this distribution. 
We then have simple histograms  with 200 bins. The results are shown in Fig. 3 for 
$Im \ \beta$ = 0.1, 0.115, 0.13 and 0.145.  
This allows us to observe the broadening of the four central peaks as $Im \ \beta$ increases. 
For instance, the two most central peaks are quite narrow up to $Im \ \beta$ = 0.1, but their 
width becomes comparable to their separation when $Im \ \beta\  > $ 0.13. One should bare in mind that 
such distributions should be understood together with the interlaced distributions for the imaginary 
part which follow similar patterns. It is clear that complex zeros found in regions where there are broad distributions are unreliable. The improvement in this situation obtained 
by using the $\beta$ independent density of states presented  above is discussed in a 
poster \cite{poster}. 
 
\begin{figure}
\begin{center}
\includegraphics[width=3.in,angle=270]{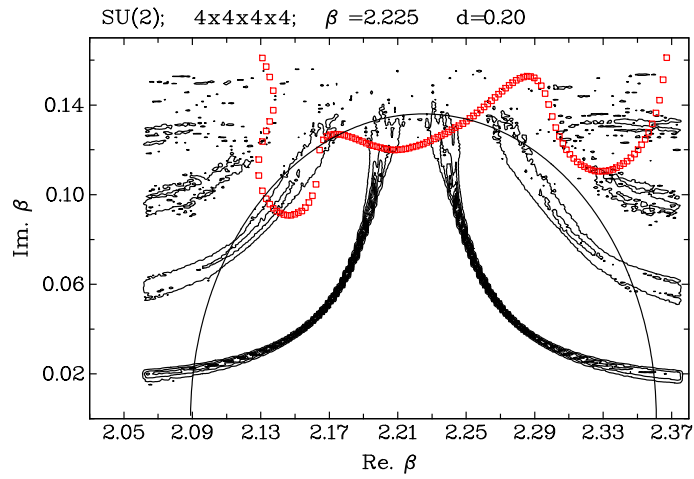}
\caption{Distribution of zeros of the real part of the partition function in the complex 
$\beta$ plane and regions of confidence described in the text.} 
\label{fig:fullcorr}
\end{center}
\end{figure}
\begin{figure}
\includegraphics[width=2.3in,angle=270]{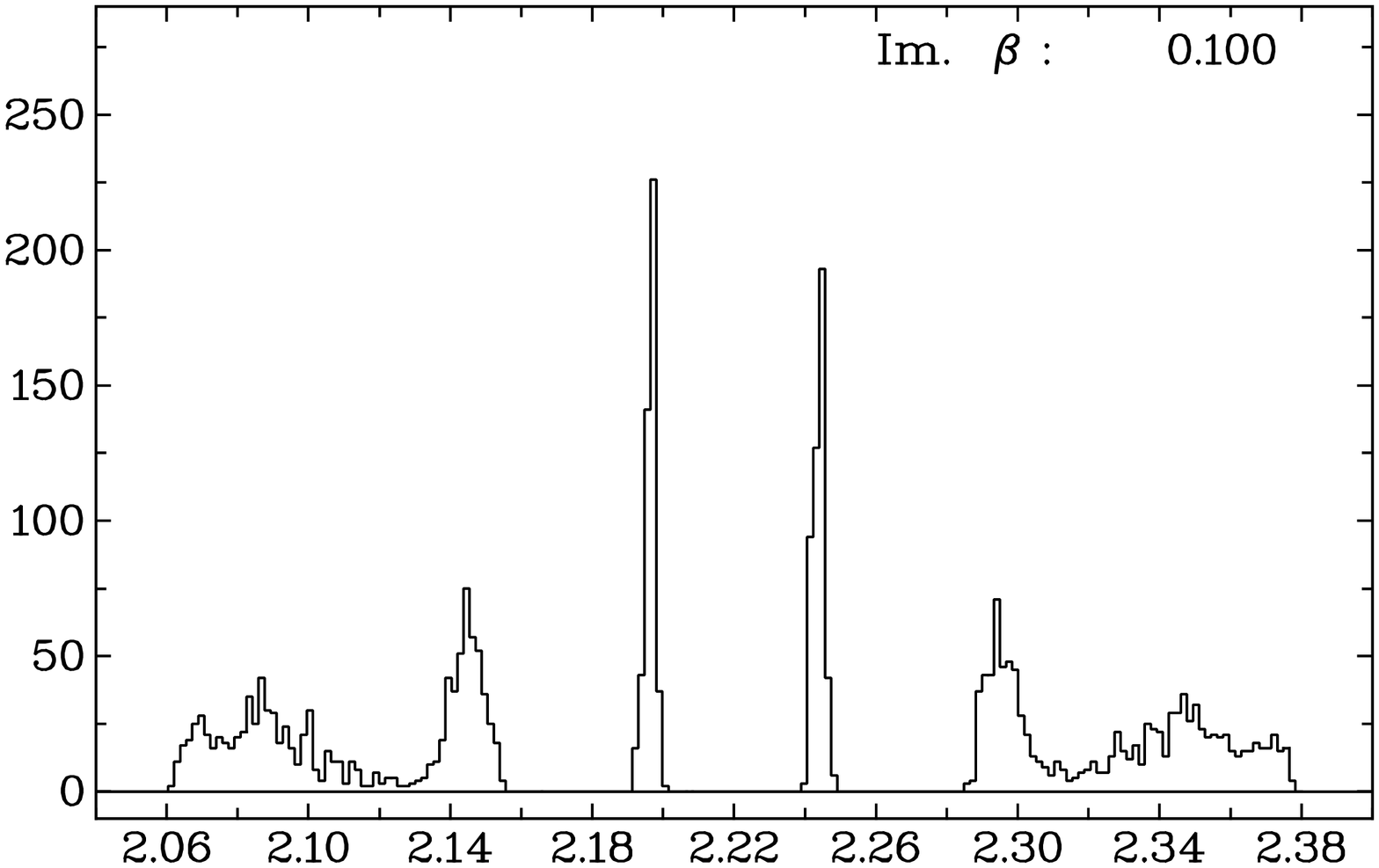}
\includegraphics[width=2.3in,angle=270]{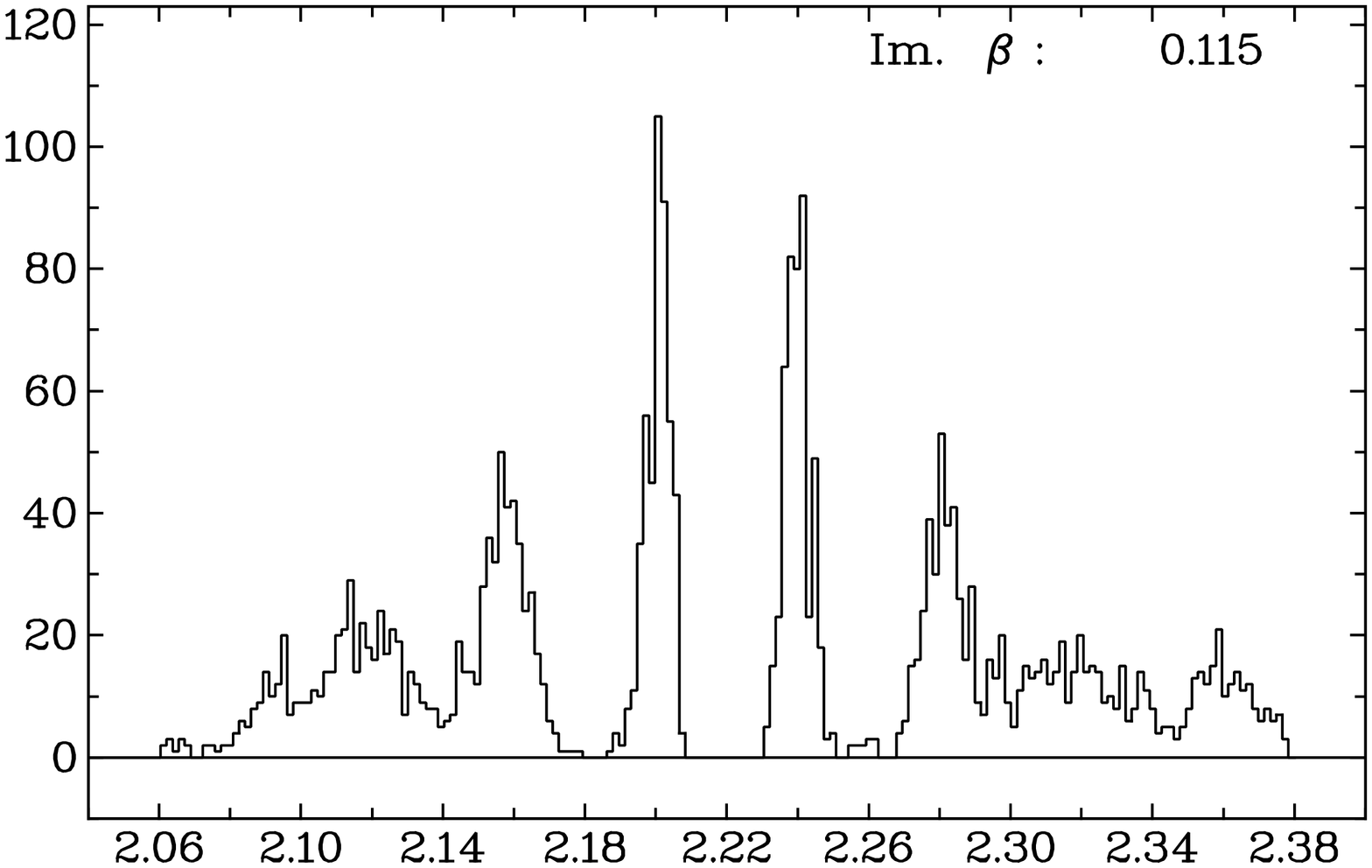}
\includegraphics[width=2.3in,angle=270]{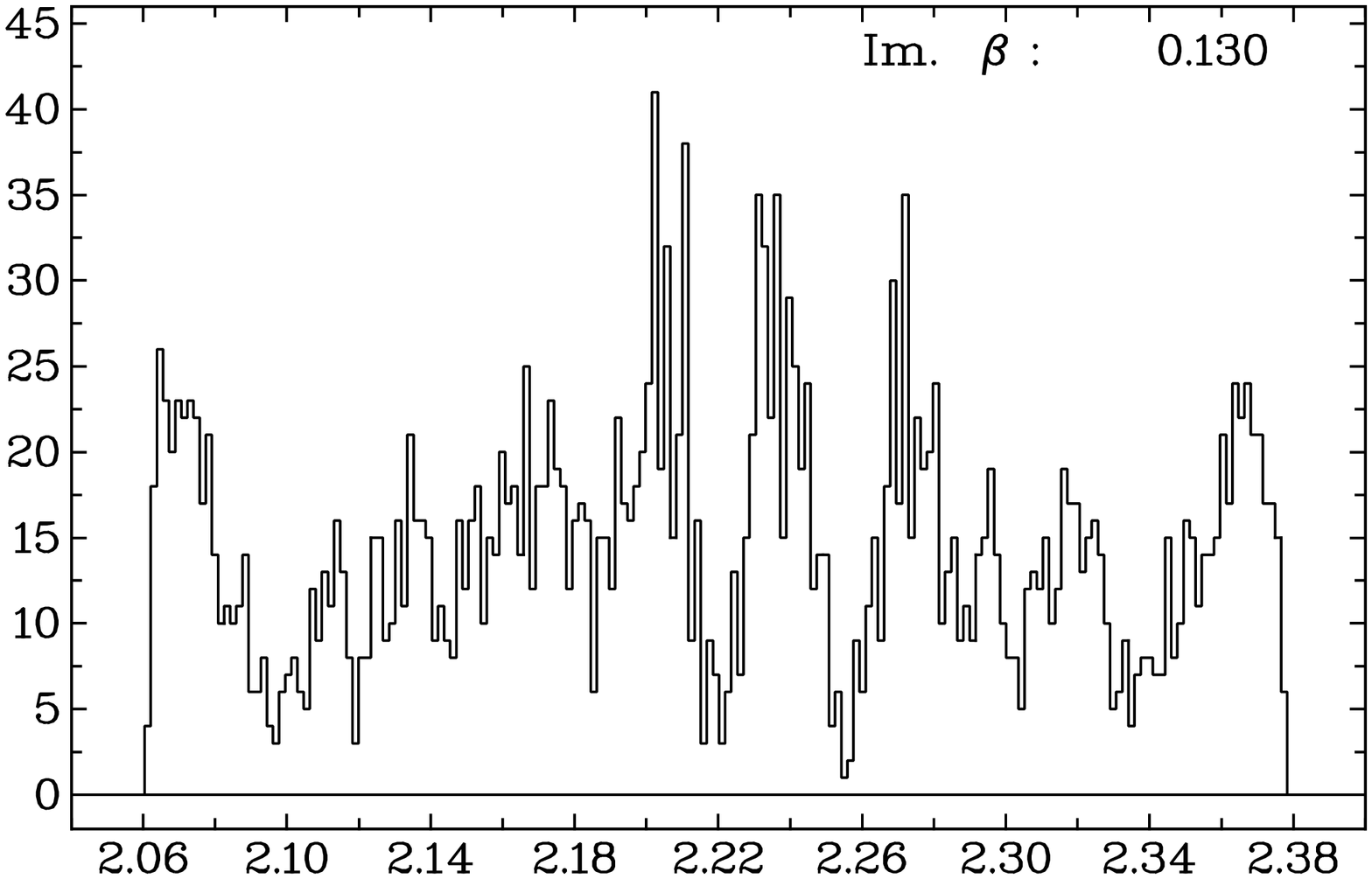}
\includegraphics[width=2.3in,angle=270]{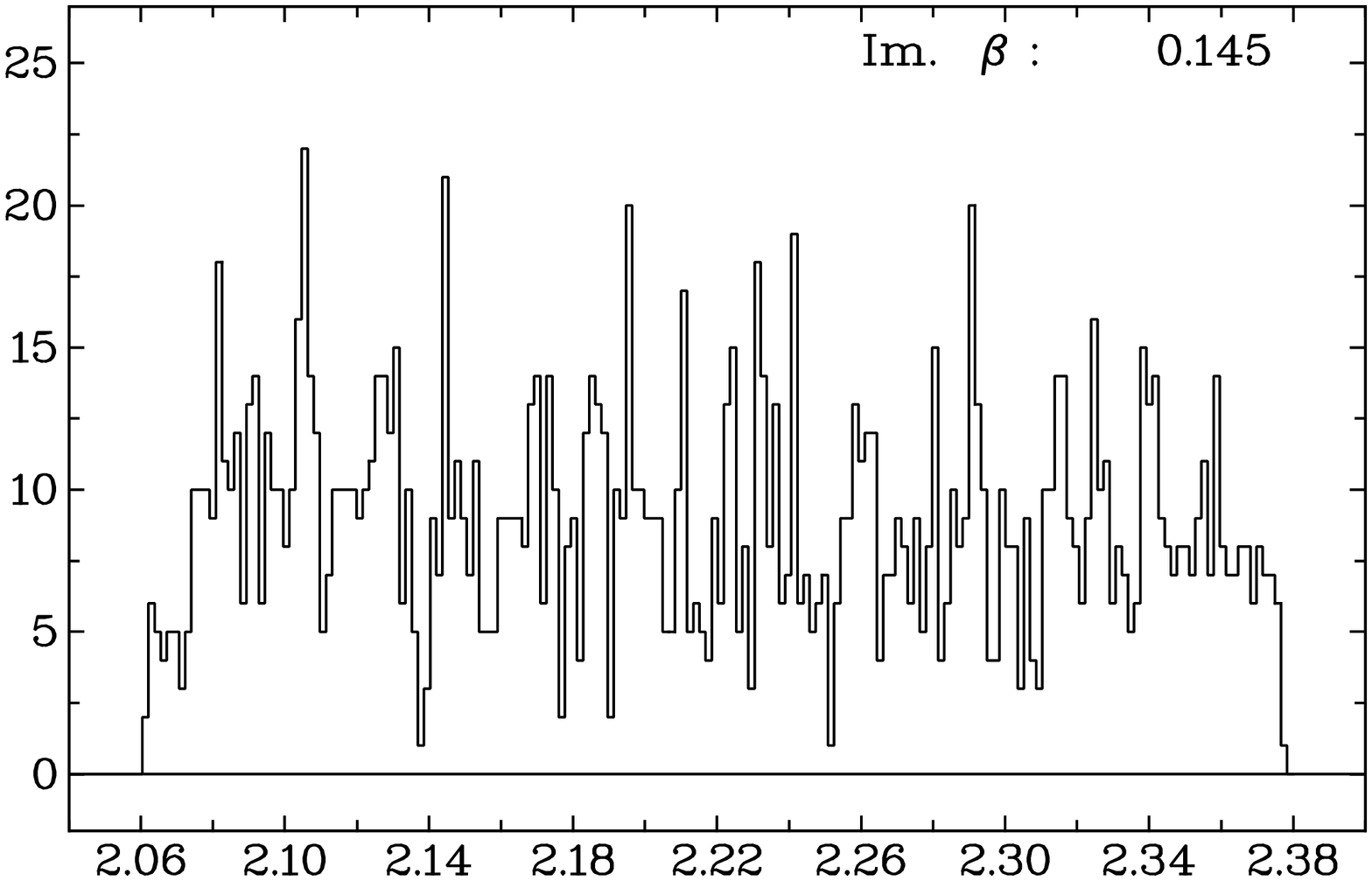}
\caption{Horizontal sections of the previous graph described in the text.}
\label{fig:corr40}
\end{figure}

\section{Conclusions}
We have calculated numerically the density of states for $SU(2)$ lattice gauge theory. 
The intermediate orders in weak and strong coupling agree well in an overlapping 
region of action values as shown in Fig. \ref{fig:wall}. However, the large order behaviors of these expansions 
appear to be similar to the corresponding ones for the plaquette. Volume effects 
can be resolved for small actions values. Corrections to the saddle point estimate need to be developed more systematically. Aprroximation of a subtracted quantity by
Chebyshev 
polynomials looks very promising. We also plan to use this method to study abelian gauge theories and the large $N$ behavior of $SU(N)$ gauge theories where interesting 
results based on the density of states have already been obtained
\cite{Bringoltz:2008zq}.

\acknowledgments

This 
research was supported in part  by the Department of Energy
under Contract No. FG02-91ER40664. 
A.V. work was supported by the Joint Theory Institute funded together by 
Argonne National Laboratory and the University of Chicago, and in part by the U.S. Department of Energy, Division of High Energy Physics and Office of Nuclear Physics, under Contract DE-AC02-06CH11357.


\begin{thebibliography}{10}

\bibitem{tomboulis07}
E.~T. Tomboulis,
\newblock {Confinement for all values of the coupling in four- dimensional
  SU(2) gauge theory}, arXiv:0707.2179
\newblock 2007.

\bibitem{tomboulis07b}
E.~T. Tomboulis,
\newblock {Deriving confinement via RG decimations},
\newblock {\em PoS}, LATTICE2007:336, 2007.

\bibitem{kogut80b}
J.~B. Kogut,
\newblock Progress in lattice gauge theory,
\newblock {\em Phys. Rept.}, 67:67, 1980.

\bibitem{third}
L.~Li and Y.~Meurice, 
\newblock About a possible 3rd order phase transition at t = 0 in 4d
  gluodynamics,
\newblock {\em Phys. Rev.}, D73:036006, 2006.

\bibitem{plaquette}
L.~Li and Y.~Meurice,
\newblock An example of optimal field cut in lattice gauge perturbation theory,
\newblock {\em Phys. Rev.}, D71:054509, 2005.

\bibitem{denbleyker:054503}
A.~Denbleyker, Daping Du, Yuzhi Liu, Y.~Meurice, and A.~Velytsky, 
\newblock Series expansions of the density of states in su(2) lattice gauge
  theory,
\newblock {\em Physical Review D (Particles and Fields)}, 78(5):054503, 2008.

\bibitem{alves90b}
Nelson~A. Alves, Bernd~A. Berg, and Sergiu Sanielevici,
\newblock Partition function zeros and the su(3) deconfining phase transition,
\newblock {\em Phys. Rev. Lett.}, 64:3107--3110, 1990.

\bibitem{alves91}
Nelson~A. Alves, Bernd~A. Berg, and Sergiu Sanielevici, 
\newblock Spectral density study of the su(3) deconfining phase transition, 
\newblock {\em Nucl. Phys.}, B376:218--252, 1992.

\bibitem{npp}
Y.~Meurice,
\newblock {The non-perturbative part of the plaquette in quenched QCD},
\newblock {\em Phys. Rev.}, D74:096005, 2006.

\bibitem{quasig}
A.~Denbleyker, D.~Du, Y.~Meurice, and A.~Velytsky,
\newblock Fisher's zeros of quasi-gaussian densities of states,
\newblock {\em Phys. Rev.}, D76:116002, 2007.

\bibitem{lat07}
A.~Denbleyker, D.~Du, Y.~Meurice, and A.~Velytsky,
\newblock {Fisher's Zeros and Perturbative Series in Gluodynamics},
\newblock {\em PoS}, LAT2007:269, 2007.

\bibitem{janke04}
W.~Janke, D.~A. Johnston, and R.~Kenna,
\newblock Phase transition strength through densities of general distributions
  of zeroes,
\newblock {\em Nucl. Phys.}, B682:618--634, 2004.

\bibitem{poster}
A.~Denbleyker, D.~Du, Y.~Liu, Y.~Meurice, and A.~Velytsky, poster presented 
at this conference.

\bibitem{gluodyn04}
L.~Li and Y.~Meurice,
\newblock Lattice gluodynamics at negative g**2,
\newblock {\em Phys. Rev. D}, 71:016008, 2005.

\bibitem{bender69}
C.~Bender and T.~T. Wu,
\newblock Anharmonic oscillator,
\newblock {\em Phys. Rev.}, 184:1231, 1969.

\bibitem{leguillou90}
J.~C. LeGuillou and J.~Zinn-Justin,
\newblock {\em Large-Order Behavior of Perturbation Theory},
\newblock North Holland, Amsterdam, 1990.

\bibitem{Bringoltz:2008zq}
Barak Bringoltz and Stephen~R. Sharpe,
\newblock {Applying the Wang-Landau Algorithm to Lattice Gauge Theory}, arXiv:0807.1275, 
\newblock 2008.

\end{thebibliography}

\end{document}